\documentclass[]{spie}  %>>> use for US letter paper
\usepackage{amsmath,amsfonts,amssymb}
\usepackage{graphicx}
\usepackage[colorlinks=true, allcolors=blue]{hyperref}
\usepackage{comment}
\usepackage{schemabloc}
\usepackage{mathtools}
\usepackage{subcaption}
\usepackage{float}

\usepackage{makeidx}
\makeindex

\usepackage[pscoord]{eso-pic}% The zero point of the coordinate systemis the lower left corner of the page (the default).

\newcommand{\placetextbox}[3]{% \placetextbox{<horizontal pos>}{<vertical pos>}{<stuff>}
  \setbox0=\hbox{#3}% Put <stuff> in a box
  \AddToShipoutPictureFG*{% Add <stuff> to current page foreground
    \put(\LenToUnit{#1\paperwidth},\LenToUnit{#2\paperheight}){\vtop{{\null}\makebox[0pt][c]{#3}}}%
  }%
}%

\title{Results from the CsI Calorimeter onboard the 2023 ComPair Balloon Flight}

\author[a]{Daniel Shy}
\author[a]{Richard S. Woolf}
\author[a]{Clio Sleator}
\author[a]{Bernard Phlips}
\author[a]{J. Eric Grove}
\author[a]{Eric A. Wulf}
\author[a]{Mary Johnson-Rambert}
\author[a]{Mitch Davis}
\author[b]{Emily Kong}
\author[c]{Thomas Caligiure}
\author[c]{A. Wilder Crosier}
\author[d]{Aleksey Bolotnikov}
\author[e,f,g]{Nicholas Cannady}
\author[d]{Gabriella A. Carini}
\author[e]{Regina Caputo}
\author[d]{Jack Fried}
\author[e, g, j]{Priyarshini Ghosh}
\author[h]{Sean Griffin}
\author[e]{Elizabeth Hays}
\author[d]{Sven Herrmann}
\author[e]{Carolyn Kierans}
\author[e,i]{Nicholas Kirschner}
\author[e,j]{Iker Liceaga-Indart}
\author[e,g,k]{Zachary Metzler}
\author[e]{Julie McEnery}
\author[e]{John Mitchell}
\author[e,g,k]{A. A. Moiseev}
\author[l]{Lucas Parker}
\author[d]{Alfred Dellapenna}
\author[e]{Jeremy S. Perkins}
\author[e,g,k]{Makoto Sasaki}
\author[e,d]{Adam J. Schoenwald}
\author[k]{Lucas D. Smith}
\author[f,e]{Janeth Valverde}
\author[j,g]{Sambid Wasti}
\author[e,f,g]{Anna Zajczyk}

\affil[a]{U.S. Naval Research Laboratory, 4555 Overlook Ave SW, Washington, DC
20375}
\affil[b]{Technology Service Corporation, Arlington, VA, 22202, USA}
\affil[c]{Naval Research Enterprise Internship Program, resident at U.S. Naval Research Laboratory, Washington, DC 20375 USA}

\affil[d]{Brookhaven National Laboratory, Upton, New York 11973, USA}
\affil[e]{NASA Goddard Space Flight Center, Greenbelt, MD, USA}
\affil[f]{Center for Space Sciences and Technology, University of Maryland, Baltimore County, 1000 Hilltop Circle, Baltimore, MD 21250, USA}
\affil[g]{Center for Research and Exploration in Space Science and Technology, NASA/GSFC, Green-
belt, MD 20771, USA}
\affil[h]{Wisconsin IceCube Particle Astrophysics Center, University of Wisconsin-Madison, 222 W Washington Ave Unit 500, Madison, WI 53703, USA}
\affil[i]{The Department of Physics, The George Washington University, 725 21st NW, Washington, DC 20052, USA}
\affil[j]{Catholic University of America, 620 Michigan Ave NE, Washington, DC 20064, USA}
\affil[k]{University of Maryland at College Park, College Park, MD 20742, USA}
\affil[l]{Space Remote Sensing and Data Science, Los Alamos National Laboratory, Los Alamos, NM 87545}

\authorinfo{Corresponding author: D. Shy, e-mail: daniel.shy.civ@us.navy.mil}
\placetextbox{0.5}{0.05}{\large\textsf{Distribution Statement A. Approved for public release: distribution is unlimited.}}%

\begin{document} 
\maketitle

\begin{abstract}

The ComPair gamma-ray telescope is a technology demonstrator for a future gamma-ray telescope called the All-sky Medium Energy Gamma-ray Observatory (AMEGO). The instrument is composed of four subsystems, a double-sided silicon strip detector, a virtual Frisch grid CdZnTe calorimeter, a CsI:Tl based calorimeter, and an anti-coincidence detector (ACD). The CsI calorimeter's goal is to measure the position and energy deposited from high-energy events. To demonstrate the technological readiness, the calorimeter has flown onboard a NASA scientific balloon as part of the GRAPE-ComPair mission and accumulated around 3 hours of float time at an altitude of 40 km. During the flight, the CsI calorimeter observed background radiation, Regener-Pfotzer Maximum, and several gamma-ray activation lines originating from aluminum.

\end{abstract}


\section{Introduction}
\label{sec:intro} 

The ComPair instrument is a technology demonstrator for the All-sky Medium Energy Gamma-ray Observatory (AMEGO)~\cite{Kierans_Amego}. It consists of 4 major detector systems: a double-sided silicon strip detector (DSSD), a virtual Frisch grid CdZnTe calorimeter, a CsI:Tl based calorimeter, and an anti-coincidence detector (ACD). We previously reported on the development of the ComPair instruments in Shy et alli (2022)~\cite{compairDev}.

\begin{figure}[H]
     \centering
     \begin{subfigure}[t]{0.45\textwidth}
         \centering
         \subcaption{ $\quad \quad$}
         \includegraphics[height=0.65\textwidth]{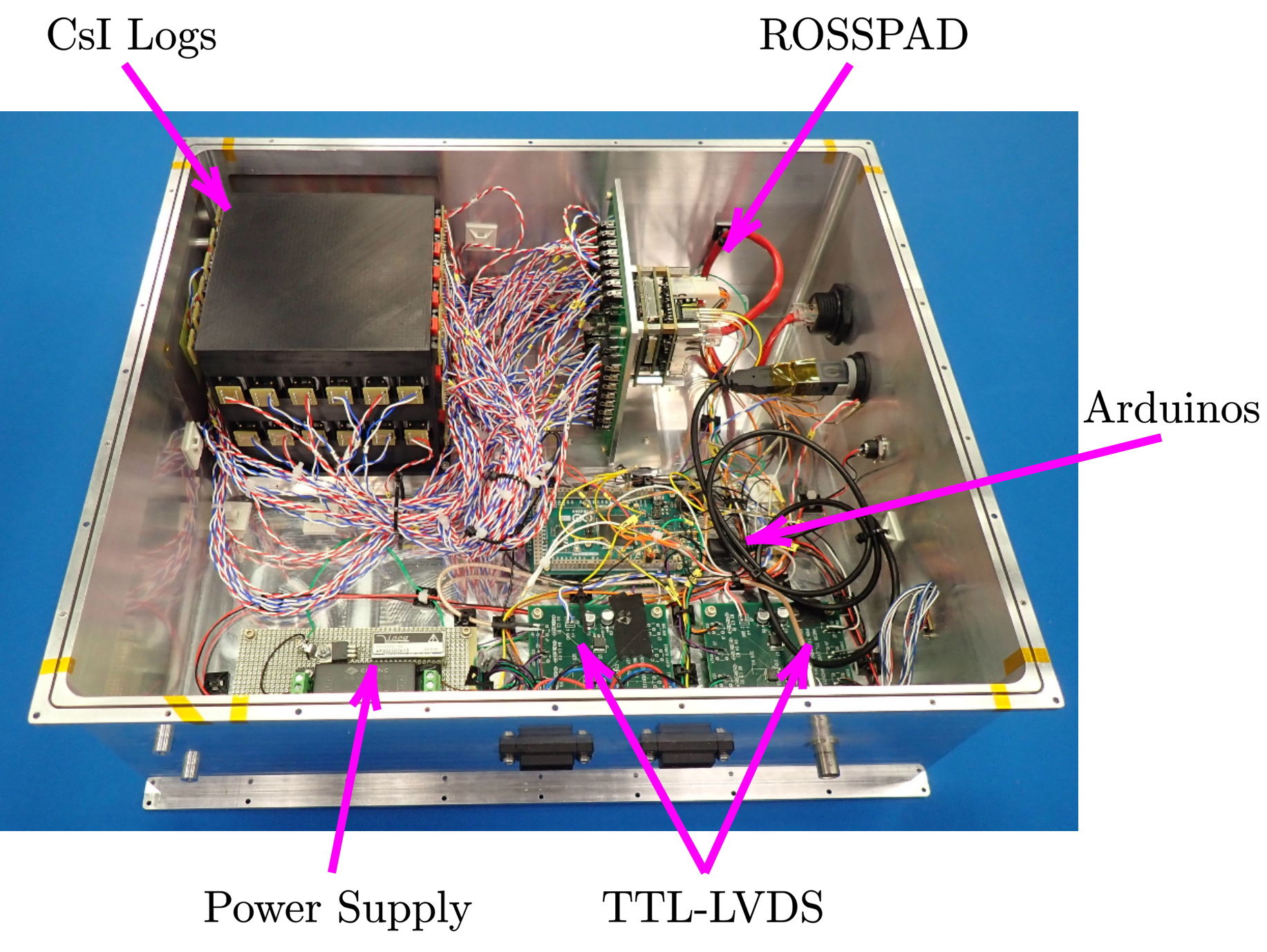}
         \label{fig:csi}
     \end{subfigure}
     \hspace{0mm}
     \begin{subfigure}[t]{0.45\textwidth}
         \centering
         \subcaption{}
         \includegraphics[height=0.65\textwidth]{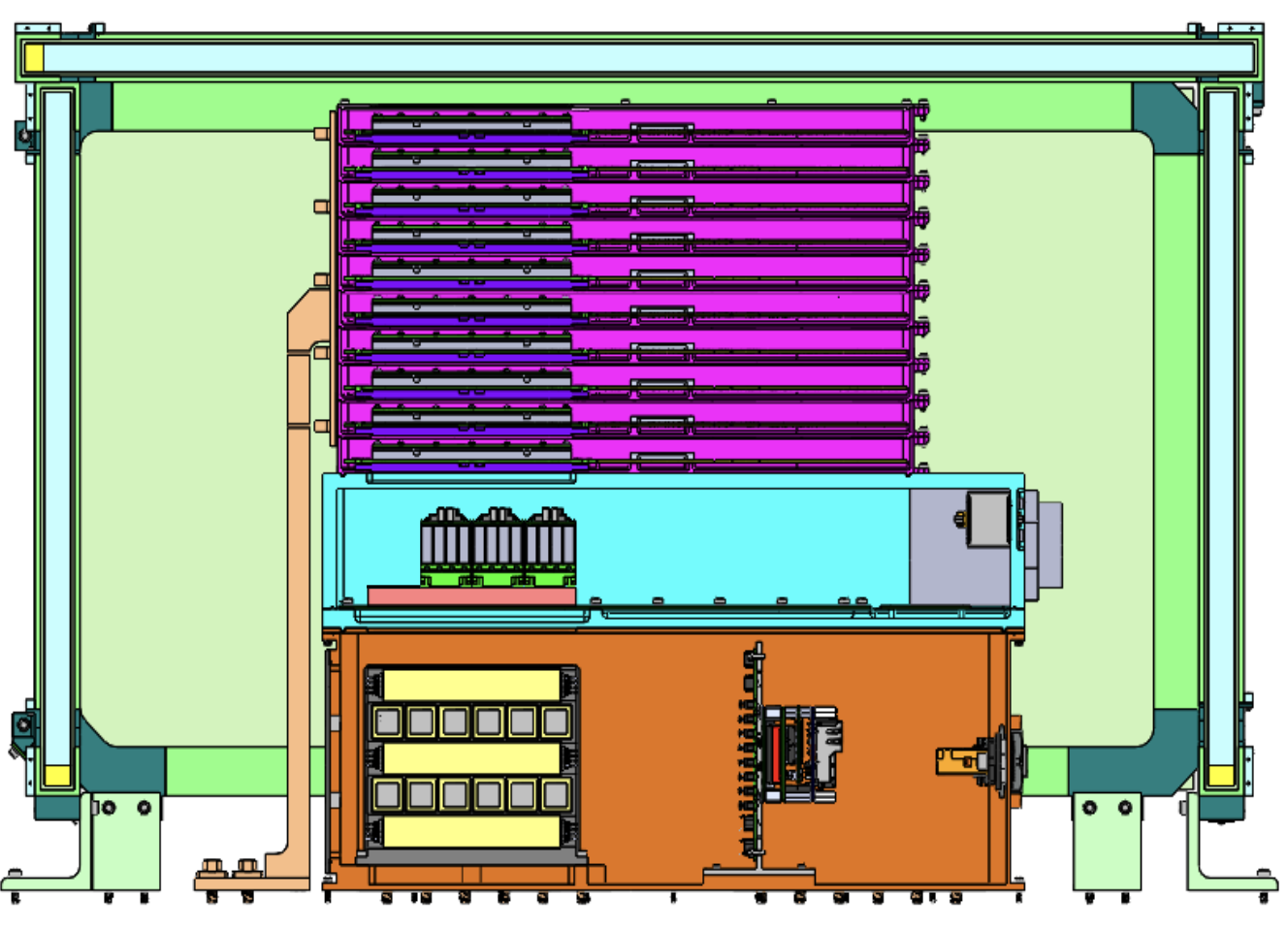}
         \label{fig:ComPair}
     \end{subfigure}

        \caption{(a) Picture of the CsI Calorimeter (image adopted from~\cite{compairDev}). (b) Cross-sectional view of the ComPair instrument.}
        \label{fig:instrumentPics}
\end{figure}

A complete description of the CsI subsystem is available in Shy et alli (2022)~\cite{compairCsI}. Briefly, the calorimeter is composed of 30 CsI:Tl scintillator logs with a volume of $1.67 \times 1.67 \times 10 \ \mathrm{cm}^3$ that each have SiPMs on each end. The SiPMs are read out by an IDE AS ROSSPAD~\cite{rosspad}, which is a front-end for SiPMs that uses four Silicon Photomultiplier Readout ASICs (SiPHRA). Each log has roughly a dynamic energy range of $250 \ \mathrm{keV}$ to $30 \ \mathrm{MeV}$. Although the ROSSPAD has not previously flown on a balloon, the SiPHRA ASIC has flown GMoDem instrument balloon flight~\cite{GMOD_Balloon}. A custom power supply that bypasses the standard ROSSPAD unit biases the 60 SiPMs required to read the calorimeter. Command and control of the ROSSPAD was accomplished via the ComPair flight computer. The flight computer makes use of a VersaLogic BayCat single-board computer that is maintained by the Los Alamos National Laboratory. Fig.~\ref{fig:csi} shows a nearly top-view of the CsI calorimeter with the lid off while Fig.~\ref{fig:ComPair} shows a side view CAD model of the entire ComPair instrument. The CsI calorimeter is located at the bottom of the stack colored orange in the figure.

To demonstrate the increase in the technological readiness level (TRL), the ComPair instrument flew onboard a NASA scientific balloon~\cite{nasaBalloon}. This manuscript focuses on the results of the CsI calorimeter. Sec.~\ref{sec:flight} briefly summarizes the balloon flight while Sec.~\ref{sec:spectra} presents results from the CsI during the balloon flight.

\section{The GRAPE-{ComPair} Balloon Flight}
\label{sec:flight} 

\begin{figure}[H]
  \centering
  \includegraphics[trim={0cm 0cm 0cm 0cm}, clip, width=0.5\linewidth]{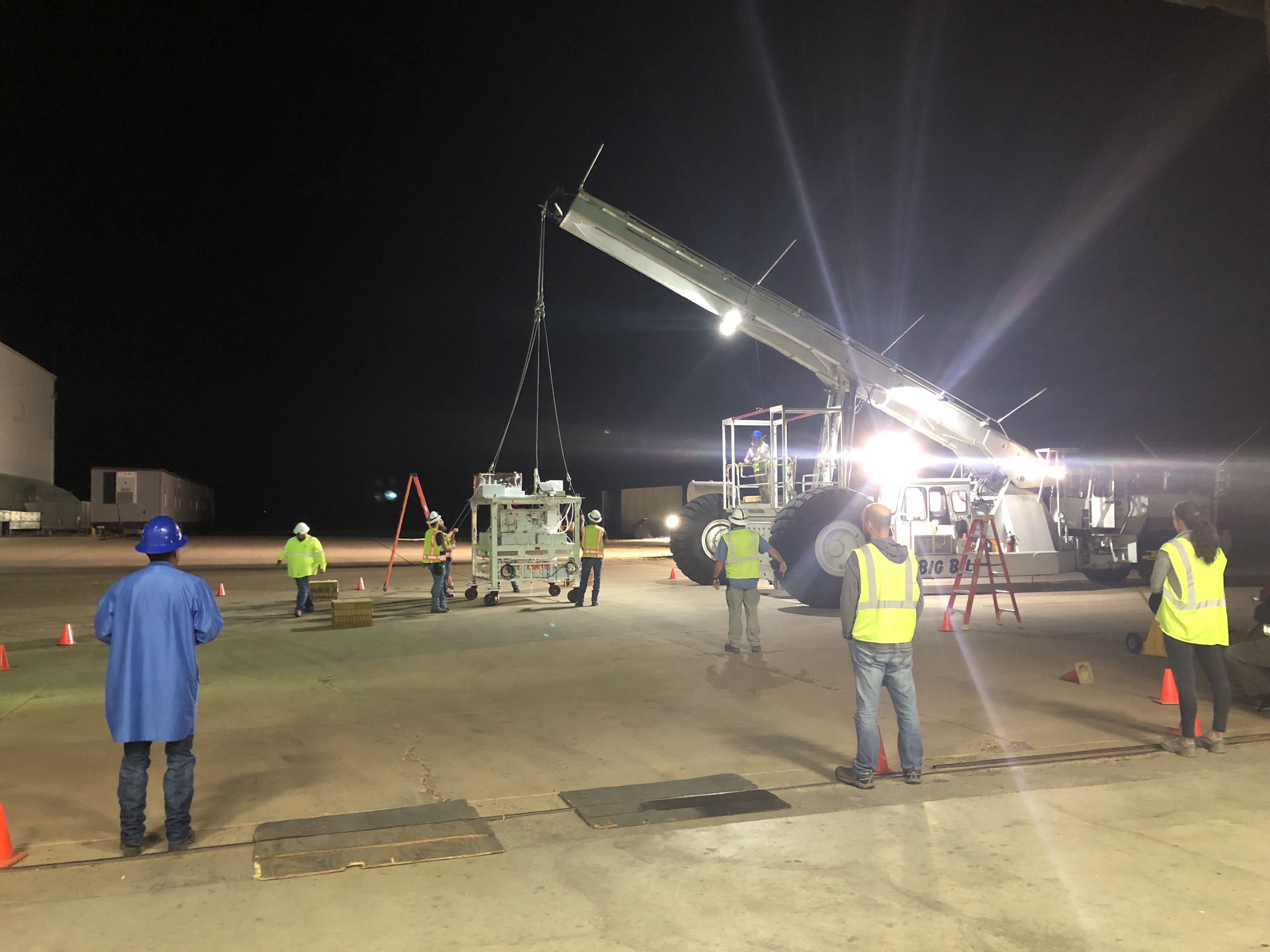}
  \caption{The GRAPE-ComPair gondola in the staging process to roll out for a launch attempt.}
  \label{fig:balloonStaging}
\end{figure}

\begin{figure}[H]
  \centering
  \includegraphics[trim={0cm 0cm 0cm 0cm}, clip, width=0.75\linewidth]{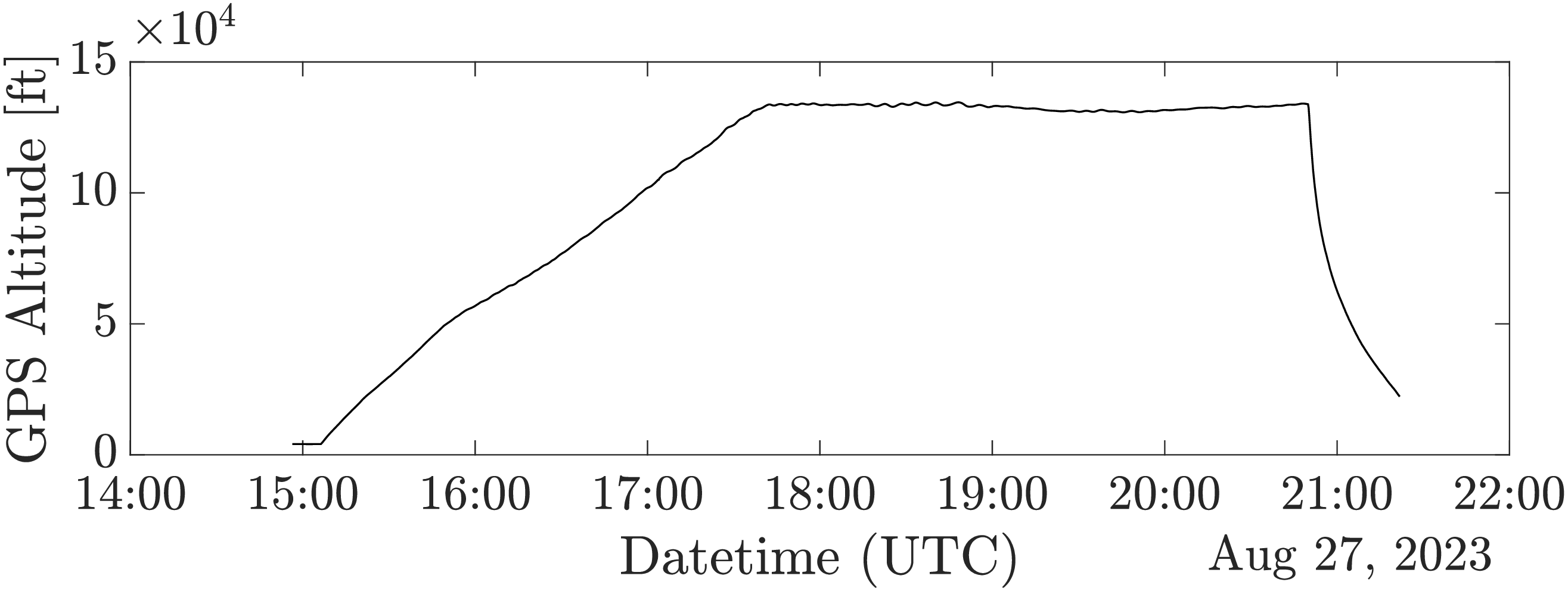}
  \caption{Flight profile of the GRAPE-ComPair balloon showing altitude as a function of time.}
  \label{fig:altVtime}
\end{figure}

Balloon launch services were provided by NASA's Columbia Scientific Balloon Facility (CSBF). The ComPair flight was accomplished as a piggyback to the GRAPE mission. Fig.~\ref{fig:balloonStaging} shows the GRAPE-ComPair gondola being staged for a flight attempt. It was launched out of Ft. Sumner, New Mexico on August 27, 2023 at 15:06 UTC. Fig.~\ref{fig:altVtime} plots the altitude and Global Positioning System (GPS) profile of the flight. The balloon climbed for $\sim$2.5 hours and then floated $\sim$3 hours at 133 kft (40 km). During the flight, the system was restarted twice due to the power distribution unit overheating, $\sim$4 hours  and $\sim$5.1 hours into the flight. The balloon landed southwest of Albuquerque, NM $\sim$6.5 hours after launch. More information on the balloon flight is available in the companion paper by L. Smith et al. titled \textit{The 2023 Balloon Flight of the ComPair Instrument}~\cite{Smith}.

\section{Calorimeter Results From the Flight}
\label{sec:spectra} 

\begin{figure}[H]
  \centering
  \includegraphics[trim={0cm 0cm 0cm 0cm}, clip, width=0.75\linewidth]{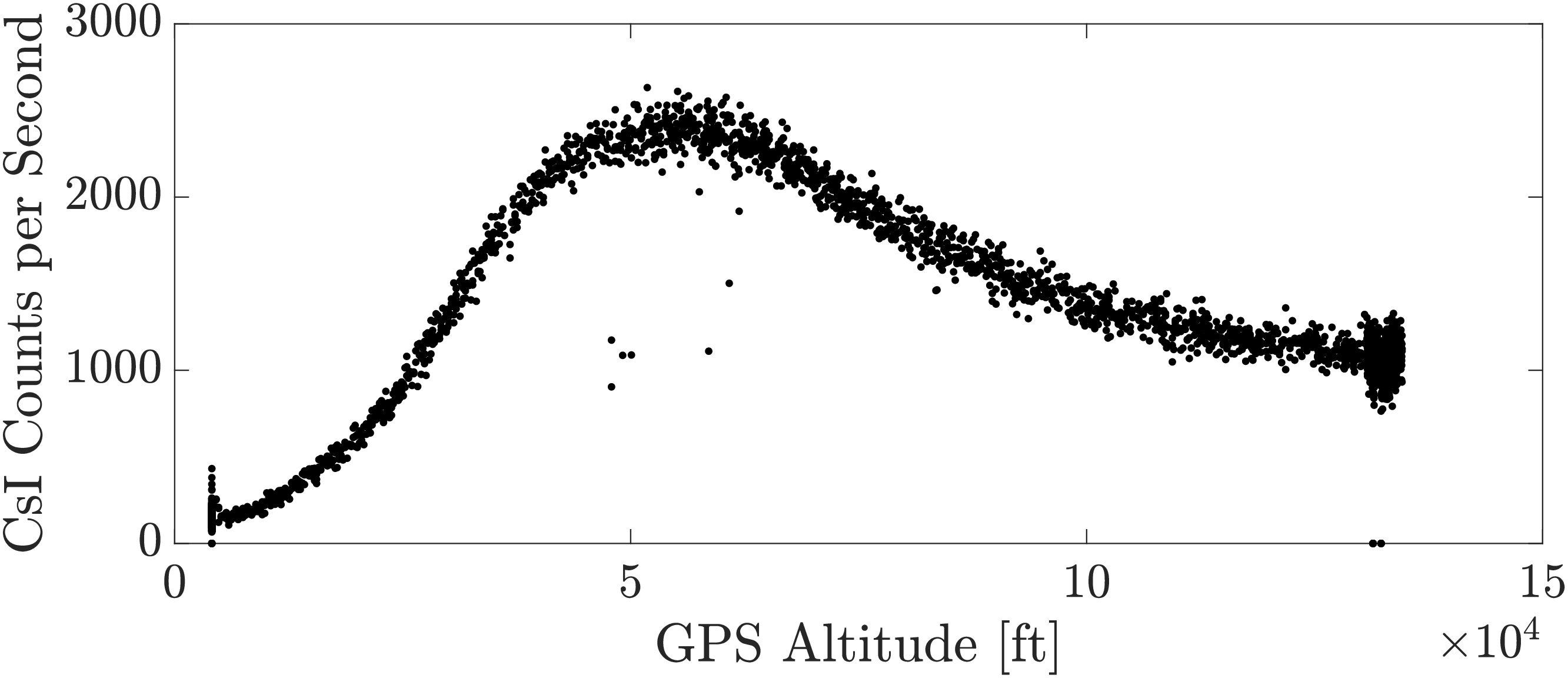}
  \caption{Count rate of the CsI calorimeter as a function of altitude.}
  \label{fig:ctsValt}
\end{figure}

The calorimeter operated in science mode throughout the flight (except during power cycles). The nominal operating mode of the calorimeter's data acquisition unit (DAQ) is flash-trigger mode, in that, if one channel is triggered, all channels will be read out. In addition, the ComPair trigger module~\cite{MakotoTM} could issue a trigger to the calorimeter, which will initiate a CsI trigger. Fig.~\ref{fig:ctsValt} plots the CsI count rate as a function of altitude. The plot starts when the instrument is hoisted by the launch vehicle, where the majority of the acquired counts are due to terrestrial gamma-ray background. Following the launch, as the balloon increased in altitude, the count rate decreased as terrestrial background radiation became negligible. However, the count rate then increased until the Regener-Pfotzer Maximum, or the region of peak radiation intensity due to the interaction of cosmic rays with the atmosphere~\cite{REGENER1935}. In this flight, it is visible around 55 kft (16.8 km) feet. The count rate would then decrease until we reach the float altitude. We set 30-minute acquisition periods, which will result in a reset of the system when the system starts writing to a new file. In addition to the data saving to onboard hard drives, housekeeping data is telemetered to the ground at a rate of 2 Hz where the health of the instrument is monitored during the flight. The housekeeping data includes the ASIC pedestal baseline and CsI count rate.

\begin{figure}[H]
  \centering
  \includegraphics[trim={0cm 0cm 0cm 0cm}, clip, width=0.75\linewidth]{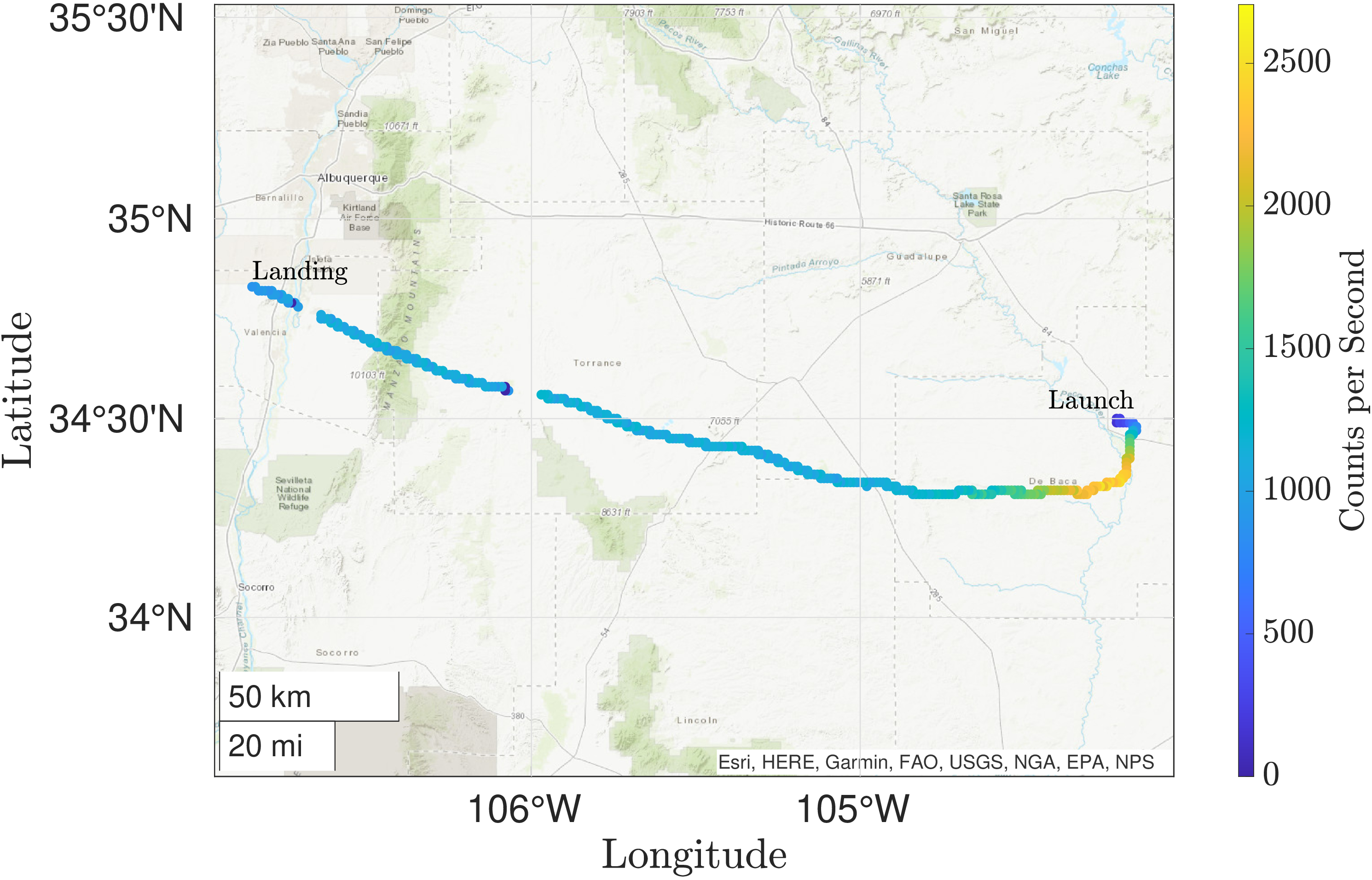}
  \caption{GPS location of the balloon with the color of the points representing the CsI calorimeter count rate at that location.}
  \label{fig:flightProfile}
\end{figure}

Fig.~\ref{fig:flightProfile} plots the GPS location with the color of the scatter point representing the count rate experienced there. The figure shows that as the balloon starts climbing, it is heading south. As it crosses Regener-Pfotzer Maximum, the balloon begins to turn westward as it clears the troposphere.

\subsection{ASIC Pedestal Response}
\label{sec:pedestal}

\begin{figure}[H]
  \centering
  \includegraphics[trim={0cm 0cm 0cm 0cm}, clip, width=0.75\linewidth]{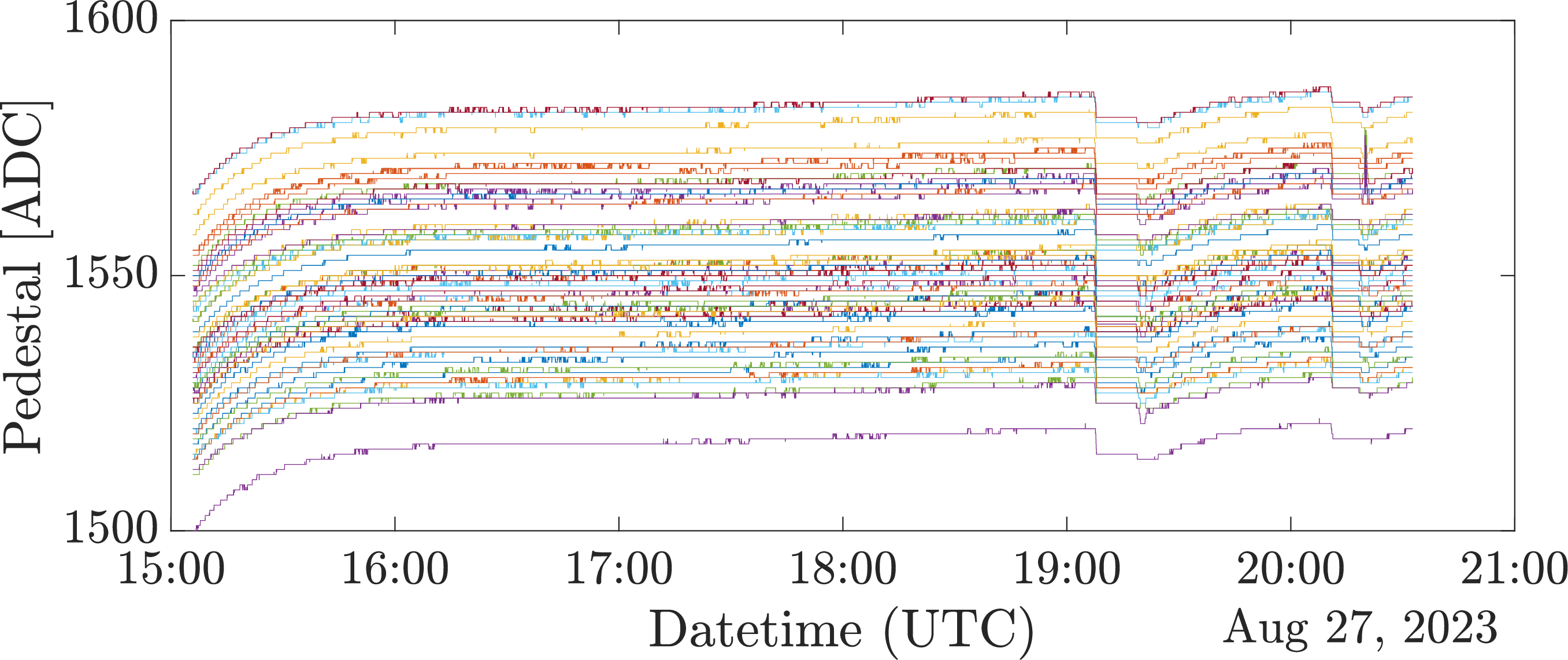}
  \caption{The ASIC pedestals associated with each of the 60 channels over time.}
  \label{fig:pedestal}
\end{figure}

The SiPHRA ASIC applies a baseline pedestal, which is an electrical offset added to each channel to prevent the value from going negative. During energy reconstruction, we subtract the pedestal's analog to digital converted (ADC) value and then scale based on the energy calibration. Fig.~\ref{fig:pedestal} plots the pedestal values for all 60 channels as a function of time. Over the course of the flight, we observe an increase in pedestal value, presumably due to the increase in temperate of the instrument, which is consistent with the trend observed in the thermal vacuum chamber (TVaC)~\cite{compairCsI}. After the power cycles, which occurred at 19:08 and 20:11, we observed a decrease in the pedestal value due to the instrument cooling down, followed by a rise again when the instrument was powered back on.

\subsection{Acquired Energy Spectrum}
\label{sec:spectrum}

\begin{figure}[H]
  \centering
  \includegraphics[trim={0cm 0cm 0cm 0cm}, clip, width=0.75\linewidth]{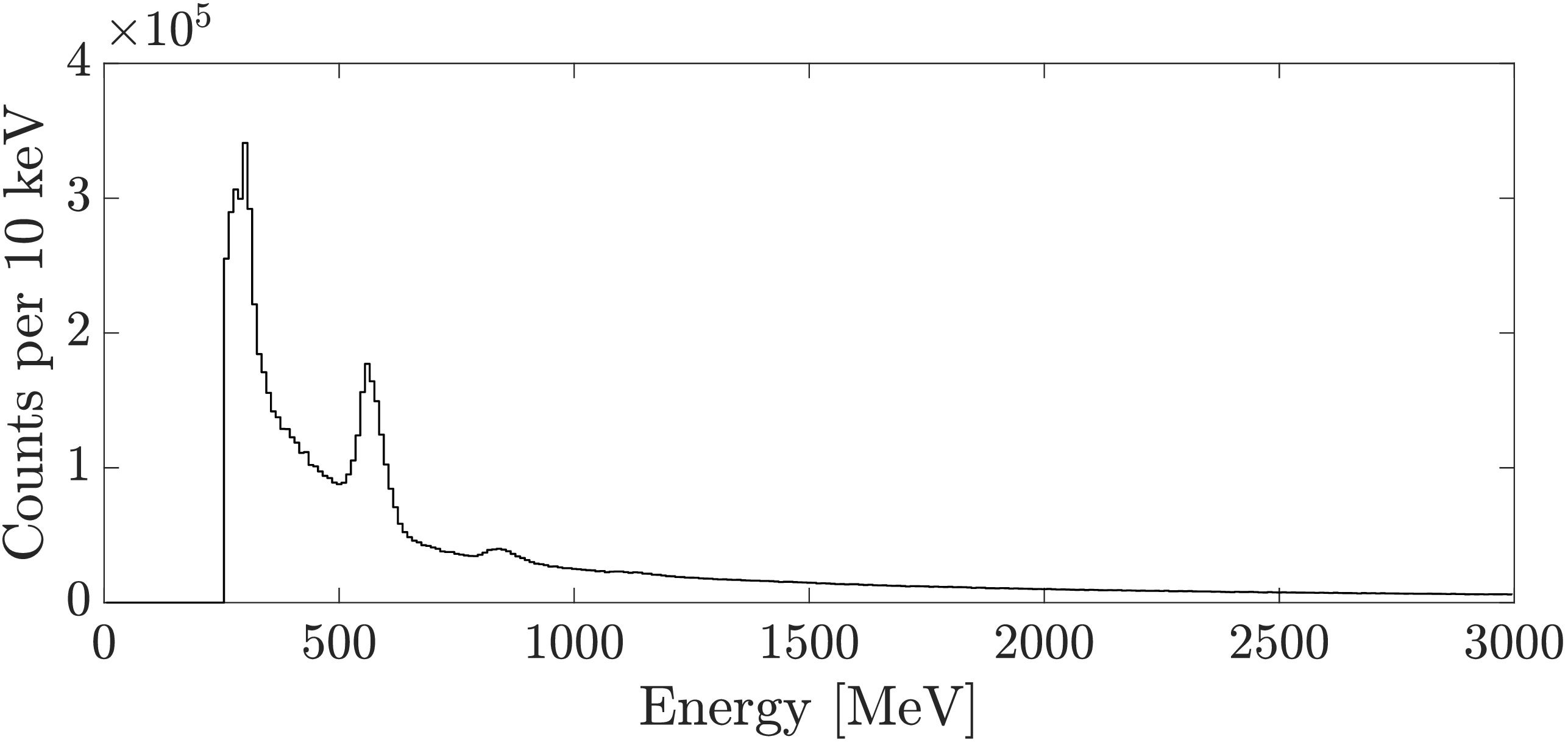}
  \caption{CsI calorimeter spectra accumulated over the entire flight.}
  \label{fig:cumSpectra}
\end{figure}

Fig.~\ref{fig:cumSpectra} shows the acquired spectrum over the entire flight (climb and float). All the energy plots in this work use a single calibration taken before the flight and are therefore subject to change due to the varying environmental conditions experienced in flight. The $511 \ \mathrm{keV}$ peak is visible in the spectrum. Also visible is a peak at $843.8 \ \mathrm{keV}$, likely due to the $^{27}\mathrm{Al}(\mathrm{n,p})^{27}\mathrm{Mg}$ reaction and decay with a branching ratio of around 72\%~\cite{alActivation}. The decay of $^{27}\mathrm{Mg}$ also releases a $1014.4 \ \mathrm{keV}$ with a branching ratio of 28\%. Only a minor feature in the spectra displays evidence of the $1014.4 \ \mathrm{keV}$ gamma. This line is more visible in the time-dependent spectra.

\begin{figure}[H]
  \centering
  \includegraphics[trim={0cm 0cm 0cm 0cm}, clip, width=0.75\linewidth]{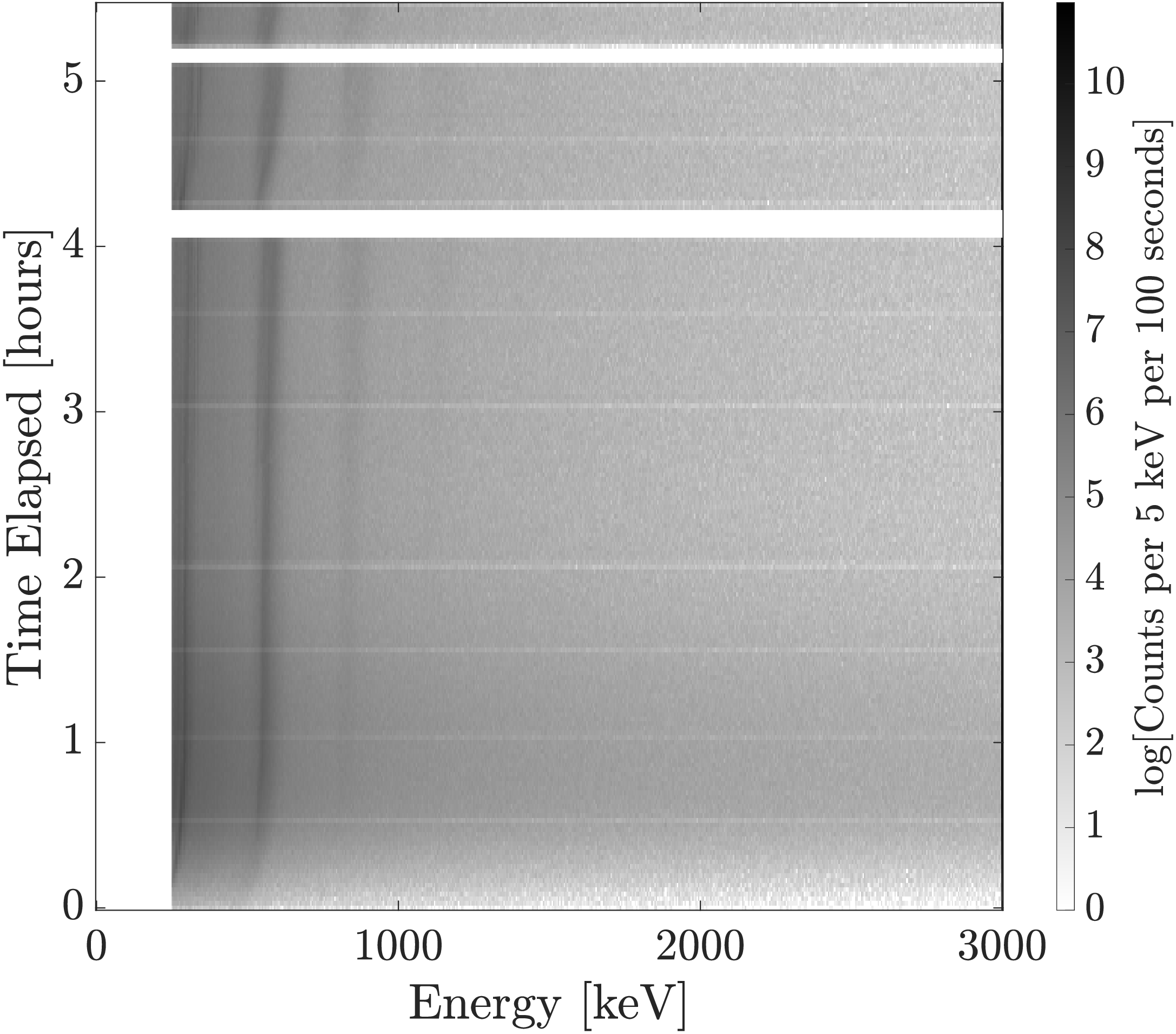}
  \caption{Energy-time response as measured during the balloon flight.}
  \label{fig:energyTime}
\end{figure}

Fig.~\ref{fig:energyTime} plots the spectrogram recorded by the CsI calorimeter or the recorded energy spectra as a function of time for 100-second chunks. The systematic horizontal lines represent the 30-minute run segmentation. The two white bands are when the instrument is not recording data due to the power outages (4 and 5.1 hours into the flight). Vertical features represent gamma-ray lines such as the $511 \ \mathrm{keV}$ and $843.8 \ \mathrm{keV}$ line. Towards the end of the flight, a hint of the $1014.4 \ \mathrm{keV}$ becomes visible.

At the beginning of the flight and immediately after power cycling, the gamma-ray lines increase in gain. This is likely due to the temperature increase experienced by the entire instrument being on and operational. The change in energy gain with temperature was also observed in TVaC~\cite{compairCsI}.

\iffalse

\subsection{Regener-Pfotzer Maximum}
\label{sec:maximum}

\begin{figure}[H]
  \centering
  \includegraphics[trim={0cm 0cm 0cm 0cm}, clip, width=0.75\linewidth]{Results/maximum/multiplicityVErg.eps}
  \caption{Max.}
  \label{fig:multiplicityVErg_max}
\end{figure}

\begin{figure}[H]
  \centering
  \includegraphics[trim={0cm 0cm 0cm 0cm}, clip, width=0.75\linewidth]{Results/maximum/spectraVNInt.eps}
  \caption{Energy Response as a function of the number of interactions.}
  \label{fig:energyNInteractions_max}
\end{figure}

\fi

\subsection{Float Altitude}
\label{sec:float} 

\begin{figure}[H]
  \centering
  \includegraphics[trim={0cm 0cm 0cm 0cm}, clip, width=0.75\linewidth]{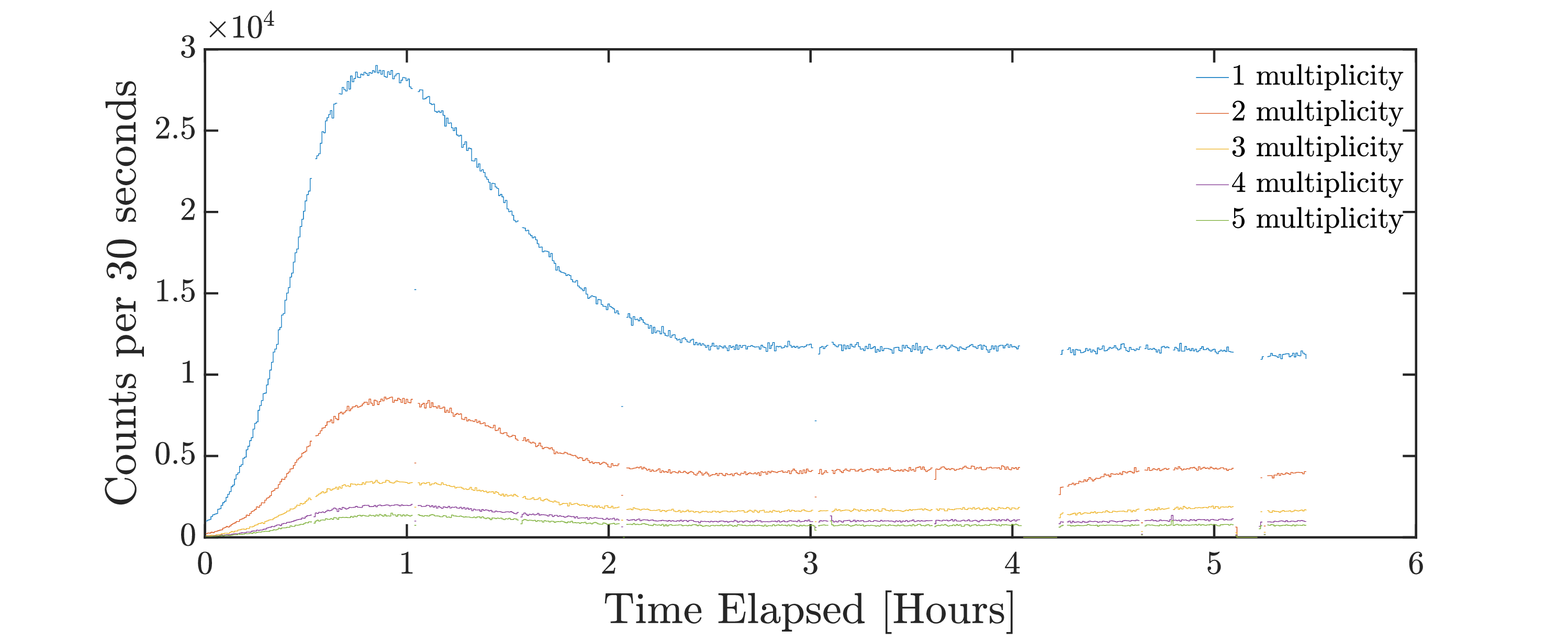}
  \caption{Count rate over the flight for different multiplicities.}
  \label{fig:cpsVsMulti}
\end{figure}

This section explores the response of the CsI calorimeter to the background radiation at the $133 \ \mathrm{kft}$ float altitude. At that altitude, we expect a large background due to protons, electrons/positrons, muons, and secondary atmospheric gamma rays~\cite{balloonBackground}. Fig.~\ref{fig:ctsValt} can be further unpacked by breaking down the recorded multiplicity over time and plotting Fig.~\ref{fig:cpsVsMulti}. We define `multiplicity' as the number of logs that recorded an interaction for a given event. We observe that right after the power outages the count rate dropped and began to rise. This is once again likely due to the temperature effects on the energy gain. As the gain increases with temperature, the energy-equivalent threshold also changes. The instrument's threshold is therefore lowered with the increase in gain, thereby increasing the count rate.

\begin{figure}[H]
  \centering
  \includegraphics[trim={0cm 0cm 0cm 0cm}, clip, width=0.75\linewidth]{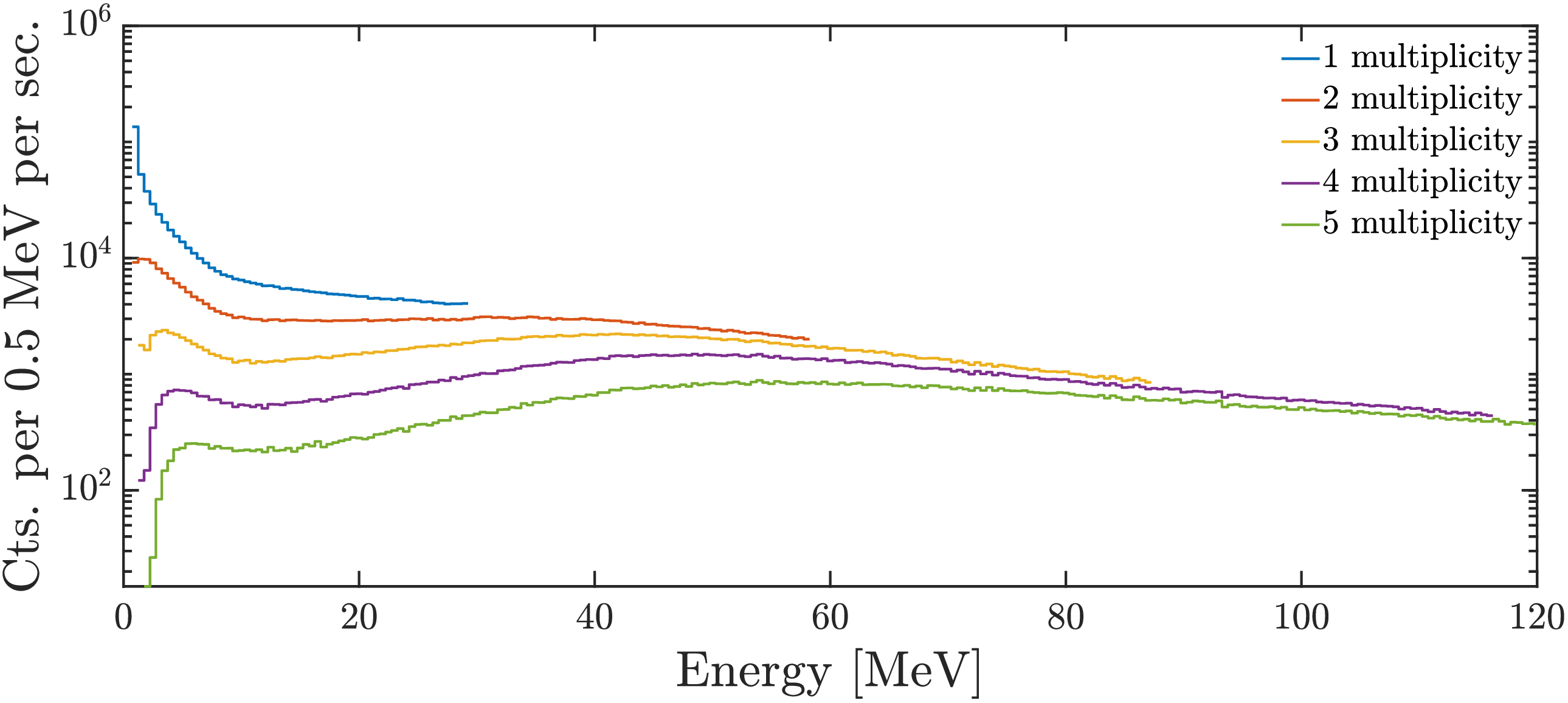}
  \caption{Energy response as a function of multiplicity at float altitude.}
  \label{fig:energyNInteractions_float}
\end{figure}

Fig.~\ref{fig:energyNInteractions_float} plots the recorded energy spectrum broken down by multiplicity. As seen in Fig.~\ref{fig:cpsVsMulti}, a multiplicity of 1 is most prominent where the gamma-ray background is the predominant species by flux~\cite{balloonBackground}. A trend is also seen that the mean of the energy spectra increases with the increase in multiplicity.

\begin{figure}[H]
  \centering
  \includegraphics[trim={0cm 0cm 0cm 0cm}, clip, width=0.75\linewidth]{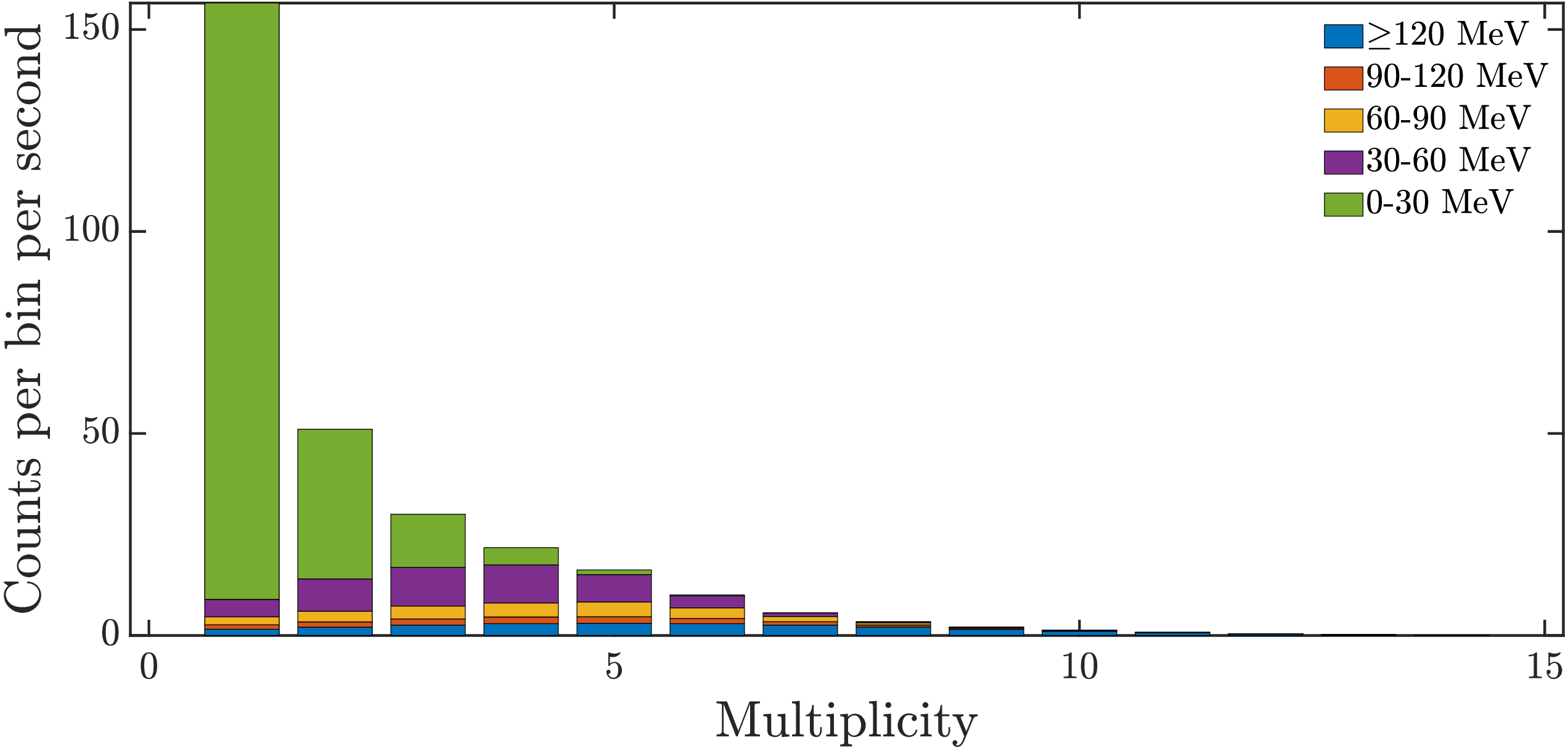}
  \caption{Multiplicity distribution as a function of energy ranges from $0$ to $120 \ \mathrm{MeV}$ at float altitude. Note that this is a stacked histogram plot.}
  \label{fig:multiplicityVErg_Float}
\end{figure}

\begin{figure}[H]
  \centering
  \includegraphics[trim={0cm 0cm 0cm 0cm}, clip, width=0.75\linewidth]{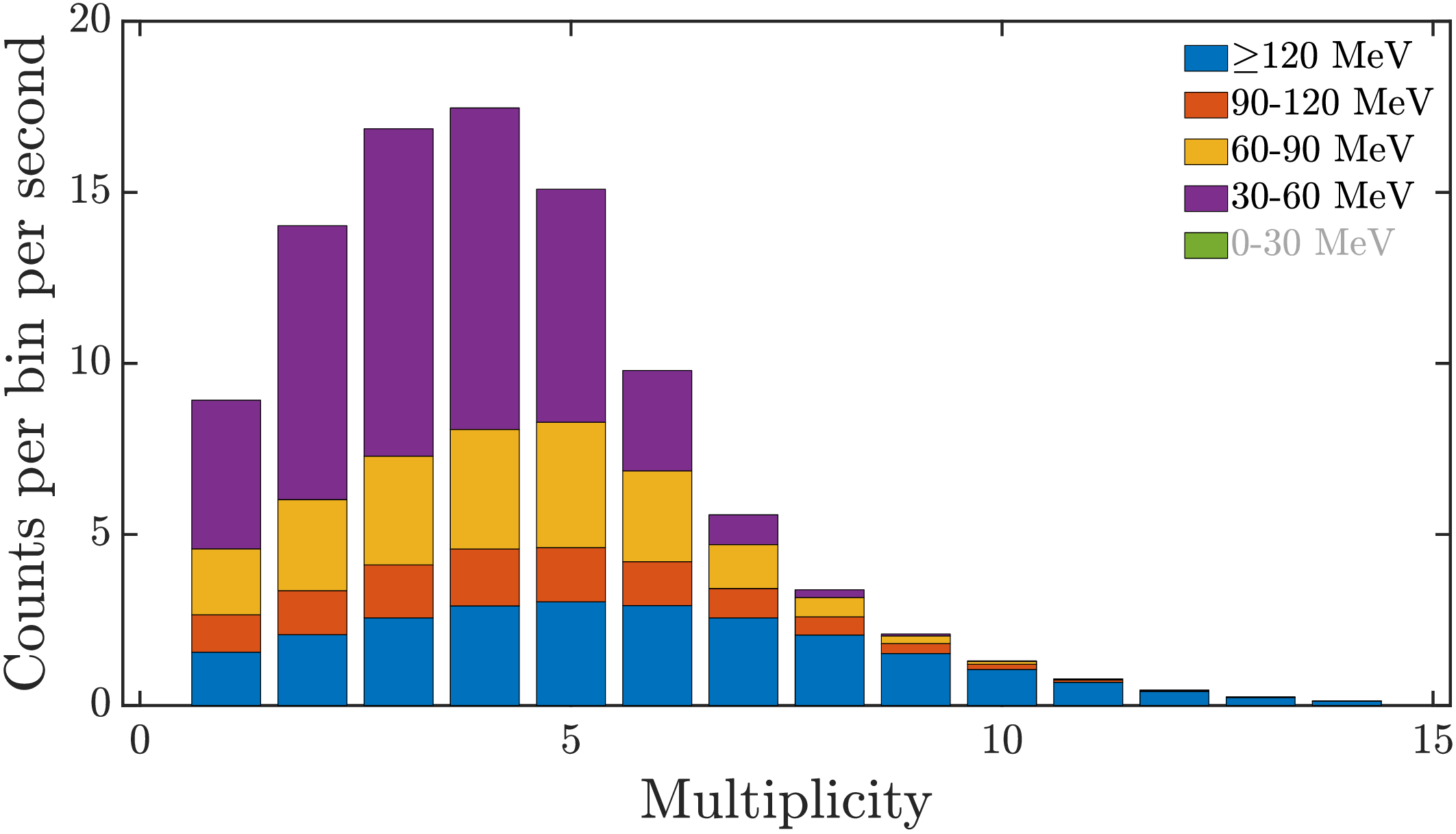}
  \caption{Multiplicity distribution as a function of energy ranges from $30$ to $120 \ \mathrm{MeV}$ at float altitude. Note that this is a stacked histogram plot.}
  \label{fig:multiplicityVErg_GreaterThan30}
\end{figure}

We can study the effect of energy on the multiplicity with Fig.~\ref{fig:multiplicityVErg_Float}, which plots the multiplicity when gating with different energy ranges. Once again, the 1 multiplicity is dominating. We see the centroid of the distribution increase with higher energies. In Fig.~\ref{fig:multiplicityVErg_GreaterThan30}, we remove the $0$ to $30 \ \mathrm{MeV}$ curve to highlight the high energy response. We see that between $30$ and $90 \ \mathrm{MeV}$, the multiplicities have a centroid between 4-5. Higher energy events naturally trigger more logs as they are possibly spallation events.

\begin{figure}[H]
  \centering
  \includegraphics[trim={0cm 0cm 0cm 0cm}, clip, width=0.75\linewidth]{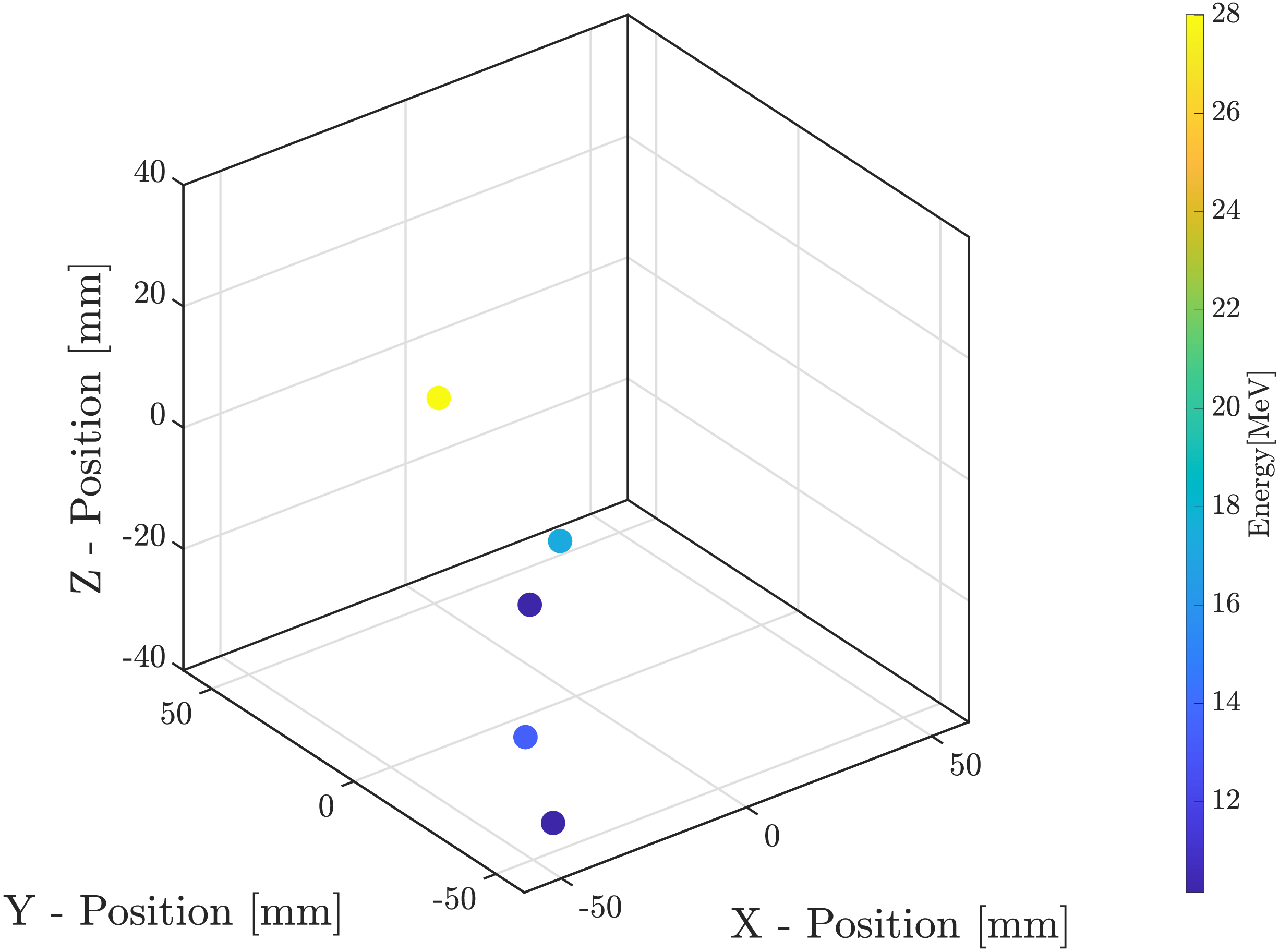}
  \caption{Scatter plot of the tracks associated with a high energy event.}
  \label{fig:3dRecon6520}
\end{figure}

\begin{figure}[H]
  \centering
  \includegraphics[trim={0cm 0cm 0cm 0cm}, clip, width=0.75\linewidth]{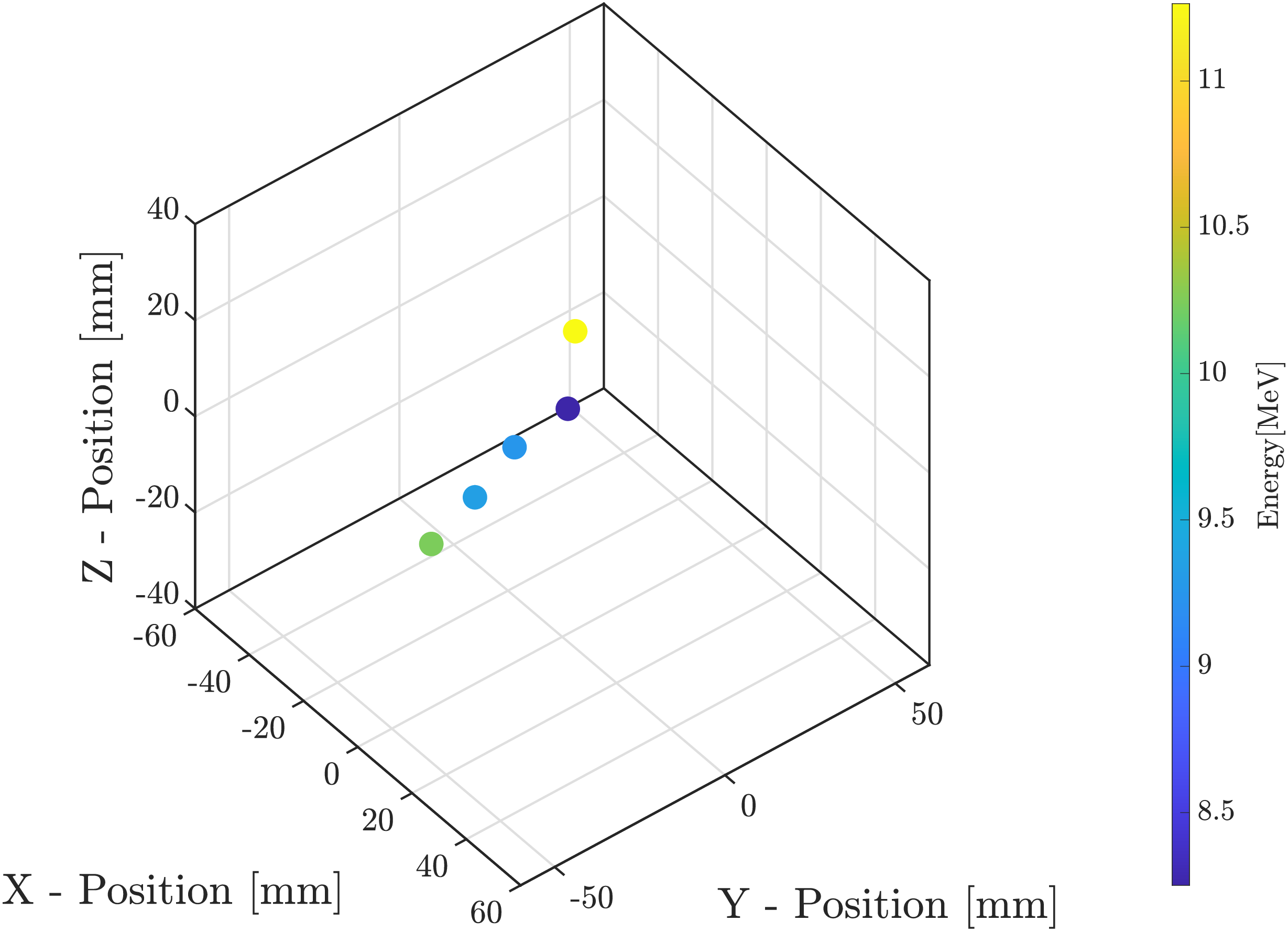}
  \caption{Scatter plot of the tracks associated with a high energy event.}
  \label{fig:3dRecon19203}
\end{figure}

Fig.~\ref{fig:3dRecon6520} and~\ref{fig:3dRecon19203} present scatter plots of tracks from high-energy events showing a single interaction in all 5 layers of the calorimeter. The color of each marker corresponds to the deposited energy at that site. Note that no error bars are added, however, the bars in the Z direction are $1.67 \ \mathrm{cm}$ thick, and depending on a layer, the errors will either be the bar thickness or the position depth error ($\sim 1 \ \mathrm{cm}$).

\section{Conclusion}
\label{sec:compair} 

The CsI calorimeter completed a scientific balloon mission as part of the ComPair balloon flight and demonstrated operations in a near-space environment. During the flight, the CsI calorimeter observed the Regener-Pfotzer Maximum and measured several gamma-ray lines consistent with aluminum activation. Development is currently underway for ComPair-2, the next iteration of the instrument to demonstrate technology for the AMEGO-X mission concept~\cite{AMEGOX}. New technology under development includes using CMOS pixel silicon sensors as the tracker layer, led by NASA Goddard Space Flight Center, and dual-gain SiPMs to read out the CsI Calorimeter, led by the U.S. Naval Research Laboratory. ComPair-2 aims to be a prototype/flight-like instrument module.

\clearpage

\acknowledgments % equivalent to \section*{ACKNOWLEDGMENTS}       

This work is supported under NASA Astrophysics Research and Analysis (APRA) grants NNH14ZDA001N-APRA, NNH15ZDA001N-APRA, NNH18ZDA001N-APRA, NNH21ZDA001N-APRA. Daniel Shy is supported by the U.S. Naval Research Laboratory's Jerome and Isabella Karle Distinguished Scholar Fellowship Program. A. W. Crosier and T. Caligiure would like to acknowledge the Office of Naval Research NREIP Program. The authors are grateful to NASA CSBF and the GRAPE team for accommodating the ComPair instrument.

% References
\bibliography{report} % bibliography data in report.bib
\bibliographystyle{spiebib} % makes bibtex use spiebib.bst

\end{document}